\documentclass [12pt]{article}
\usepackage{amssymb}
\usepackage{amsmath}
\usepackage[dvips]{epsfig}
\begin{document}
\begin{center}
\Large\bf{Radiation-enhanced diffusion of impurity atoms in
silicon layers }
\\[2ex]
\normalsize
\end{center}

\begin{center}
O. I. Velichko
\\
%

{\it E-mail address (Oleg Velichko):} velichkomail@gmail.com
\end{center}
\bigskip

\textbf{Abstract } Modeling of the phosphorus radiation-enhanced
diffusion in the course of implantation of high-energy protons
into an elevated-temperature silicon substrate and during its
treatment in a hydrogen-containing plasma with addition of a
diffusant has been carried out. It follows from the results
obtained that the radiation-enhanced diffusion occurs by means of
formation, migration, and dissociation of ``impurity atom --
silicon self-interstitial'' pairs being in a local thermodynamic
equilibrium with substitutionally dissolved impurity atoms and
nonequilibrium point defects generated due to external
irradiation.

The resulting value of the average migration length of
nonequilibrium silicon self-interstitials decreases from 0.19 $\mu
$m for proton energy of 140 keV to 0.09 and 0.08 $\mu $m for
energies of 110 and 80 keV, respectively. The decrease of the
average migration length with the proton energy can be due to the
interaction of silicon self-interstitials with the vacancies
generated at the surface or with the defects formed in the
phosphorus implanted region.

Based on the pair diffusion mechanism, a theoretical investigation
of the form of impurity profiles that can be created in thin
silicon layers due to the radiation-enhanced diffusion was carried
out. It is shown that depletion of the uniformly doped silicon
layer occurs during plasma treatment except for the silicon --
insulator interface where a narrow region with a high impurity
concentration is formed. The results of calculations give a clear
evidence in favor of further investigation of various doping
processes based on the radiation-enhanced diffusion, especially
the processes of plasma doping, to develop a cheap method for the
formation of strictly assigned impurity distributions in the local
semiconductor domains.

\section{Introduction}
An important trend in the present-day electronics is the
diminishing of the dimensions of semiconductor devices
\cite{Packan-00}. Now, the active regions of advanced integrated
microcircuits and optoelectronics devices can reach geometric
dimensions of about 10 nm. Another trend is the use of diverse
multilayer structures to achieve the required parameters of
devices. The multilayer structures are employed in the devices
based on compound semiconductors, silicon or germanium. For
example, much attention has been paid recently to engineering
Ge$_{1-x}$Si$_{x}$ layers
\cite{Lim-00,Portavoce-04,Berbezier-09,Kasper-12} and to the
silicon on insulator technology (SOI)
\cite{Langdo-04,Hamilton-07,Bhandari-08}. The implementation of
these trends means that interfaces will exert a significant
influence on the dopant--defect system and, consequently, on the
electrophysical parameters of devices. For example, the interfaces
present themselves as planar defects and can act as a source
\cite{Pichler-04} or sink of point defects \cite{Lamrani-04}.
Thus, distributions of point defects can be nonuniform in the
vicinity of the interface. Nonuniform distributions of
nonequilibrium point defects can also be formed due to ion
implantation that is widely used for the production of modern
microcircuits and other electronic devices \cite{ITRS-11}. If the
concentration of point defects generated due to ion implantation
or ion bombardment exceeds the thermally equilibrium
concentration, the radiation-enhanced diffusion (RED) occurs.

The present analysis of the up-to-date trends in the modern
technology of electronic devices allows the formulation of the
main idea of the proposed theoretical investigation. The goal of
this work is to describe the main features of a coupled evolution
of dopant--defect system in thin silicon layers resulting from the
nonuniformity of point defect distributions formed due to the
surface influence or due to ion implantation (bombardment). The
expected results will be important for the design of devices as
well as for the semiconductor physics and technology.

\section{Models of the radiation-enhanced diffusion}

The equation describing the impurity diffusion due to the
formation, migration, and dissociation of the ``impurity atom --
vacancy'' and ``impurity atom -- silicon self-interstitial'' pairs
has been obtained earlier in \cite{Velichko-84}. It was supposed
that nonuniform distributions of point defects, including defects
in the neutral charge state, can be formed and that the mass
action law is valid for the substitutionally dissolved impurity
atoms, nonequilibrium self-interstitials or vacancies, and the
pairs. The equation takes account of different charge states of
all mobile and immobile species and the drift of the pairs and
point defects in a built-in electric field, although only the
concentrations of neutral point defects are involved in an
explicit form. In the case of low impurity concentration $C\le
n_{i} $, the equation obtained can be presented in the form

\begin{equation} \label{DifEqLow}
\frac{\partial \, C}{\partial \, t} =D_{i}^{E} \, \Delta \,
\left(\tilde{C}^{V\times } C\right)+D_{i}^{F} \, \Delta \,
\left(\tilde{C}^{I\times } C\right) .
\end{equation}

Here $C$ is the concentration of substitutionally dissolved
impurity atoms, $n_{i}$ is the intrinsic carrier concentration,
$D_{i}^{E}$ and $D_{i}^{F}$ are the diffusivities of impurity
atoms in intrinsic silicon due to the ``impurity atom -- vacancy''
and ``impurity atom -- silicon self-interstitial'' pairs,
respectively, $\tilde{C}^{V\times}$ and $\tilde{C}^{I\times}$ are
the concentrations of vacancies and silicon interstitial atoms in
the neutral charge state normalized to the thermally equilibrium
concentrations of these species, $C_{eq}^{V\times}$ and
$C_{eq}^{I\times}$, respectively. It is worth noting that
$\tilde{C}^{V\times}=\tilde{C}^{V}$ and $\tilde{C}^{I\times}
=\tilde{C}^{I}$ for the low impurity concentration $C\le n_{i}$.
Here $\tilde{C}^{V}$ and $\tilde{C}^{I}$ are respectively the
total concentrations of vacancies and silicon interstitial atoms
normalized to the thermally equilibrium concentrations of these
species, $C_{eq}^{V}$ and $C_{eq}^{I}$.

For a number of doping processes only one kind of defect is
involved in diffusion of the main fraction of impurity atoms
\cite{Pichler-04}. Then, Eq. \eqref{DifEqLow} can be presented in
a simplified form:

\begin{equation}\label{DifEq1}
\frac{\partial \, C}{\partial \, t} =D_{i} \, \Delta \, \,
\left(\tilde{C}^{\times } C\right) ,
\end{equation}

\noindent where  is the concentration of the neutral defects
responsible for impurity diffusion normalized to the thermally
equilibrium concentration of this species, .

The equation obtained retains the basic character of the original
equations presented in \cite{Velichko-84}, namely, the ability to
describe segregation of impurity atoms, including ``uphill''
impurity diffusion. Indeed, we shall present Eq. \eqref{DifEq1} in
the following form:

\begin{equation}\label{DifEq1Drift}
\frac{\partial \, C}{\partial \, t} =D_{i} \, \nabla
\left(\tilde{C}^{\times } \nabla C\right)+D_{i} \, \nabla
\left[\left(\, \nabla \tilde{C}^{\times } \right) C\right].
\end{equation}

It is clearly seen from Eq. \eqref{DifEq1Drift} that, depending on
the gradient of the concentration of neutral point defects, the
drift term is added to the right hand side of the equation, which
has the type of the Fick second law. It means that an additional
drift flux of impurity atoms proportional to the gradient of
concentration of point defects in a neutral charge state is added
to the flux caused by a gradient of the impurity concentration. If
these fluxes are opposite-directed, a component of ``uphill''
diffusion is included to the general impurity flux. It means that
segregation of impurity atoms can be observed at great values of
the gradient of point defects in a neutral charge state instead of
the leveling effect for nonuniform impurity distribution. If the
``uphill'' diffusion component exceeds the usual diffusion flux,
described by the Fick first law, an unusual form of impurity
profile is observed. Therefore, by modeling the impurity diffusion
under conditions of nonuniform distribution of point defects one
can obtain very useful information on the microscopic mechanisms
underlying the impurity atoms and point defects migration.

In one dimension, Eqs. \eqref{DifEq1} and \eqref{DifEq1Drift} have
the following form:

\begin{equation}\label{DifEq1x}
\frac{\partial \, C}{\partial \, t} =D_{i} \, \frac{\partial ^{\,
2} \left(\tilde{C}^{\times } C\right)}{\partial \, x^{2} }
\end{equation}

\begin{equation}\label{DifEq1Driftx}
\frac{\partial \, C}{\partial \, t} =D_{i} \, \frac{\partial
}{\partial \, x} \left(\tilde{C}^{\times } \frac{\partial \,
C}{\partial \, x} \right)+D_{i} \, \frac{\partial }{\partial \, x}
\left(\frac{\partial \, \tilde{C}^{\times } }{\partial \, x}
 \, C\right) .
\end{equation}

It is worth noting that there is no difference in the mathematical
description of impurity transport processes occurring due to
equilibrium ``impurity atom -- silicon self-interstitial'' pairs
and due to the kick-out mechanism if the impurity interstitials
are in local thermodynamic equilibrium with the substitutional
impurity atoms and nonequilibrium silicon self-interstitials
\cite{Robinson-92,Velichko-08}. At the same time, there is an
essential difference between the mathematical descriptions of
impurity transport caused by the equilibrium impurity-vacancy
pairs and by the simple vacancy mechanism of diffusion when the
impurity atom and neighboring vacancy exchange places. Indeed, the
equation of impurity diffusion due to such a simple vacancy
mechanism has been derived in \cite{Uskov-72,Morikawa-80}:

\begin{equation}\label{DifEqVacMech}
\frac{\partial \, C}{\partial \, t} =D_{i} \, \frac{\partial
}{\partial \, x} \left(\tilde{C}^{V} \frac{\partial \, C}{\partial
\, x} \right)-D_{i} \, \frac{\partial }{\partial \, x}
\left(\frac{\partial \, \tilde{C}^{V} }{\partial \, x} \, C\right)
.
\end{equation}

It can be seen from this equation that the second term in the
right-hand side of it has the ``minus'' sign, whereas the second
term in the right-hand side of Eq. \eqref{DifEq1Driftx} has the
``plus'' sign. It means that the flux caused by the vacancy
gradient in the simple vacancy mechanism has an opposite direction
in comparison with the similar flux for the case of impurity
diffusion due to impurity-vacancy pairs.

\section{Preliminary studies of the radiation-enhanced \,  \,  \,  diffusion}

It follows from the analysis of Eqs. \eqref{DifEq1Driftx} and
\eqref{DifEqVacMech} that investigation of impurity redistribution
under conditions of nonuniform distribution of the point defects
responsible for the impurity diffusion allows one to draw a
conclusion regarding the character of the diffusion mechanism. For
example, by simulating the impurity redistribution investigated in
\cite{Akutagawa-79}, it was shown in \cite{Velichko-81} that the
radiation-enhanced diffusion of boron during proton bombardment of
a silicon substrate having an elevated temperature occurs due to
the equilibrium ``impurity atom -- intrinsic point defect'' pairs.
It follows from \cite{Krause-Rehberg-00,Peeva-04} that the
vacancy-type defects are formed in the region between the surface
and the average projective range of implanted ions $R_{p} $ (more
exactly, in the vicinity of ${R_{p}
\mathord{\left/{\vphantom{R_{p}
2}}\right.\kern-\nulldelimiterspace} 2} $), whereas the
interstitial stacking faults are formed near $R_{p} $ and deeper
inside. For a proton energy of 140 keV used for boron
redistribution in \cite{Akutagawa-79}, $R_{p} $ = 1.235 $\mu $m
and $R_{m} $ = 1.27 $\mu $m \cite{Burenkov-80}. Here $R_{p} $ is
the average projective range of implanted ions and $R_{m} $ is the
position of the maximum for concentration profile of implanted
protons described by the type IV Pearson distribution
\cite{Burenkov-80}. Taking into account the experimental results
of \cite{Krause-Rehberg-00,Peeva-04}, one can conclude that
silicon self-interstitials are the point defects responsible for
the RED of boron atoms, because the local minimum of impurity
concentration profile (both experimental and theoretical) lies
near$R_{m} $.

We have a more complicated situation for the radiation-enhanced
diffusion of phosphorus atoms. Indeed, in
\cite{Velichko-97,Velichko-02} simulation of phosphorus
redistribution during proton bombardment of doped silicon
substrates was carried out. The experimental data of
\cite{Akutagawa-79} were used for comparison. In the experiments
of Akutagawa et al. \cite{Akutagawa-79} the initial phosphorus
profile in the (111) oriented silicon substrates was formed due to
ion implantation with an energy of 140 keV and a dose of
1.5$\times$10$^{12}$ ions/cm$^{2}$. The choice of a very low dose
was necessary for phosphorus profiling by the differential C-V
technique to avoid the effects caused by avalanche breakdown.
Simultaneously, the effects caused by the concentration-dependent
diffusion were avoided. After the implantation and before the
proton-enhanced diffusion, the wafers were annealed at a
temperature of 750 $^{\circ}$C for 30 min in a purified argon
ambient for electrical activation of the implanted impurity. After
the enhanced diffusion, the samples were left in the target
chamber at 700 $^{\circ}$C for times longer than about 30 min for
postannealing treatment, which is expected to produce full
electrical activity of the impurity and remove the residual
radiation damage. To provide the radiation-enhanced diffusion, the
silicon substrates were implanted with 80-keV protons at a beam
density of 1.0 $\mu $A/cm$^{2}$. Proton bombardment was carried
out at a temperature of 700 $^{\circ}$C for 3, 10, 30, and 120
minutes. It was concluded from the simulation results that the RED
of phosphorus atoms is governed by the ``impurity atom --
intrinsic point defect'' pairs that are in local equilibrium with
the substitutionally dissolved phosphorus and generated point
defects. In contrast to the boron enhanced diffusion, it was found
that the maximum of point defect generation rate as well as the
local minimum of the impurity concentration profile are located
near $R_{mD} $, where $R_{mD} $ is the position of the maximum of
the energy deposited in atomic collisions in silicon under proton
bombardment \cite{Burenkov-85}. As follows from
\cite{Burenkov-85}, the value of $R_{mD} $ = 0.6782 $\mu $m is
less than the value of $R_{m} $ = 0.7922 $\mu $m. Taking into
account \cite{Krause-Rehberg-00,Peeva-04}, one can conclude that
the vacancies are rather the point defects responsible for the RED
of phosphorus, whereas nonequilibrium silicon self-interstitials
provide boron diffusion. Unfortunately, in
\cite{Velichko-97,Velichko-02} an approximate analytical solution
was used to calculate the point defect distribution. Therefore, it
is reasonable to investigate the radiation-enhanced diffusion of
phosphorus atoms more thoroughly.

\section{Simulation of phosphorus radiation-enhanced diffusion}

\subsection{Diffusion due to the exchange of places between an
impurity atom and a neighboring vacancy}

Let us first exclude the simple vacancy mechanism of diffusion
when the exchange of places between an impurity atom and a
neighboring vacancy occurs. For this purpose we can use the
experimental data of Strack \cite{Strack-63} that were obtained
for the case of a long-term treatment when the sputtering of the
surface of a semiconductor plays a significant role and,
therefore, the distribution of impurity atoms becomes stationary.
In the experiments of \cite{Strack-63} the
\textbf{\textit{p}}-type silicon with a conductivity of 200
$\Omega \,$cm was used. The treatment temperature was equal to 820
$^{\circ}$C and the rate of surface sputtering was equal
approximately to 5.28$\times$10$^{-4}$ $\mu $m/s. The impurity
distribution profile was found by removing thin layers from the
surface of a sample and measuring their sheet resistance. The
thermal diffusivity of phosphorus at the above-mentioned
temperature is equal to 1.919$\times$10$^{-8}$ $\mu $m$^{2}$/s
\cite{Haddara-00}, the intrinsic carrier concentration, to $n_{i}$
= 2.76$\times$10$^{6}$ $\mu $m$^{-3}$.

In order to calculate the impurity distribution measured in
\cite{Strack-63}, it is reasonable to introduce a new coordinate
system, $x=x^{*}-\mathrm{v_{S}t}$, bound to the moving surface of
the semiconductor and solve the diffusion equation in this mobile
coordinate system. Here $x^{*}$ is the coordinate measured from
the initial position of the semiconductor surface,
$\mathrm{v_{S}}$ is the projection of the surface velocity on the
immobile axis $x^{*}$ ($\mathrm{v_{S}}>0$ in the case of
sputtering of the surface).

Then, the equation of impurity diffusion \eqref{DifEqVacMech} in
the moving coordinate system takes the following form:

\begin{equation}\label{DifEqVacMechMov}
\frac{\partial \, C}{\partial \, t} =D_{i} \, \frac{\partial
}{\partial \, x} \left(\tilde{C}^{V} \frac{\partial \, C}{\partial
\, x} \right)-D_{i} \, \frac{\partial }{\partial \, x}
\left(\frac{\partial \, \tilde{C}^{V} }{\partial \, x}
C\right)+{\rm v}_{{\rm S}} \frac{\partial \, C}{\partial \, x} \,
.
\end{equation}

When the impurity atoms entering from the plasma are compensated
by those removed from the semiconductor surface due to sputtering,
the distributions of the impurity atoms and vacancies become
stationary in the moving system of coordinates. Then, Eq.
\eqref{DifEqVacMechMov} will be transformed into a stationary
diffusion equation:

\begin{equation}\label{DifEqVacMechStationary}
D_{i} \, \frac{d}{d\, x} \left(\tilde{C}^{V} \frac{d\, C}{d\, x}
\right)-D_{i} \, \frac{d}{d\, x} \left(\frac{d\, \tilde{C}^{V}
}{d\, x} C\right)+{\rm v}_{{\rm S}} \frac{d\, C}{d\, x} =0 \, .
\end{equation}

The ordinary differential equation \eqref{DifEqVacMechStationary}
can be solved analytically. Let us obtain such an analytical
solution for the following Dirichlet boundary conditions:

\begin{equation}\label{BounCondDir}
C(0)=C_{S} \, , \qquad \qquad \qquad C(+\infty )=0  \, ,
\end{equation}

\noindent where $C_{S}$ is the impurity concentration on the
semiconductor surface.

To obtain an analytical solution of the boundary-value problem
formulated, let us present Eq. \eqref{DifEqVacMechStationary} in
the following form:

\begin{equation}\label{DifEqStatM}
\frac{d\, }{d\, x} \left[D_{i} \, \tilde{C}^{V} \frac{d\, C}{d\,
x} -D_{i} \, \left(\frac{d\, \tilde{C}^{V} }{d\, x} \right)C+{\rm
v}_{{\rm S}} \, C\right]=0
\end{equation}

\noindent and integrate \eqref{DifEqStatM} from $x$   to $+\infty$
to obtain

\begin{equation}\label{Integration}
\left. D_{i} \, \tilde{C}^{V} \frac{d\, C}{d\, x} \,
\right|_{x}^{+\infty } -\left. D_{i} \, \left(\frac{d\,
\tilde{C}^{V} }{d\, x} \right)C\, \right|_{x}^{+\infty } +\left.
{\rm v}_{{\rm S}} C\, \right|_{x}^{+\infty } =0 \, .
\end{equation}

As soon as both the impurity and vacancies concentrations as well
as their fluxes are equal to zero on the right boundary $x=+\infty
$, the ordinary differential equation can be obtained from Eq.
\eqref{Integration}:

\begin{equation}\label{OrdinaryDifEq}
D_{i} \, \tilde{C}^{V} \frac{d\, C}{d\, x} -D_{i} \,
\left(\frac{d\, \tilde{C}^{V} }{d\, x} \right)\, C\, +{\rm
v}_{{\rm S}} \, C=0 \, .
\end{equation}

This equation can be solved using the separation of variables:

\begin{equation}\label{Separation}
\frac{d\, C\left(x\right)}{C\left(x\right)} =\frac{\displaystyle
\frac{d\, \tilde{C}^{V} \left(x\right)}{d\, x} -\frac{{\rm v}_{S}
}{D_{i} } }{\tilde{C}^{V} \left(x\right)} \, d\, x \, .
\end{equation}

Let us integrate Eq. \eqref{Separation} from 0 to $x$:

\begin{equation}\label{IntegralLim}
\left. \ln \left[\, C\left(x\right)\right]\, \, \right|_{0}^{x}
=\left. \ln \left[\, \tilde{C}^{V} \left(x\right)\right]\, \,
\right|_{0}^{x} -\frac{{\rm v}_{S} \, }{D_{i} \, } \int
_{0}^{x}\frac{1}{\tilde{C}^{V} \left(x\right)} \, d\, x \, ,
\end{equation}

\noindent or

\begin{equation}\label{Integral}
\ln \, \left[\frac{\, C\left(x\right)}{C_{S} } \right]=\ln \,
\left[\frac{\, \tilde{C}^{V} \left(x\right)}{\tilde{C}_{S}^{V} }
\right]-\frac{{\rm v}_{S} \, }{D_{i} \, } \int
_{0}^{x}\frac{1}{\tilde{C}^{V} \left(x\right)} \, d\, x \, ,
\end{equation}

\noindent where $\tilde{C}_{S}^{V}$   is the normalized
concentration of vacancies on the semiconductor surface.

Using exponentiation of Eq. \eqref{Integral}, we obtain the
expression for the distribution of impurity concentration in the
form

\begin{equation}\label{GenSolution}
C\left(x\right)=C_{S} \frac{\, \tilde{C}^{V}
\left(x\right)}{\tilde{C}_{S}^{V} } \exp \, \left[-\frac{{\rm
v}_{{\rm S}} \, }{D_{i} \, } \int _{0}^{x}\frac{1}{\tilde{C}^{V}
\left(x\right)} \, d\, x \right] \, .
\end{equation}

It follows from expression \eqref{GenSolution} that for ${\rm
v}_{{\rm S}} \ge 0$ the concentration of impurity atoms
$C\left(x\right)\to 0$ at $\tilde{C}^{V} \left(x\right)\to 0$. Let
us consider the widespread case of defect generation on the
semiconductor surface. Then, the spatial distribution of vacancies
can be described by the expression

\begin{equation}\label{ExpSolution}
\tilde{C}^{V} \left(x\right)=\tilde{C}_{S}^{V} \, \exp \,
\left[-\frac{x}{l_{i}^{V} } \right] \, ,
\end{equation}

\noindent where $l_{i}^{V} =\sqrt{d_{i}^{V} \tau _{i}^{V} \, } $
is the average migration length of nonequilibrium point defects.
Here $d_{i}^{V} $ and $\tau _{i}^{V} $ are the diffusivity and
average lifetime of vacancies in intrinsic silicon, respectively.

Substituting \eqref{ExpSolution} into \eqref{GenSolution} and
calculating the integral obtained, we find that the impurity
distribution is determined by the expression

\begin{equation}\label{PartialSolution}
C\left(x\right)=C_{S} \, \frac{\, \tilde{C}^{V}
\left(x\right)}{\tilde{C}_{S}^{V} } \exp \, \left\{-\frac{{\rm
v}_{S} \, l_{i}^{V} }{D_{i} \, \tilde{C}_{S}^{V} } \left[\exp
\left(\frac{x}{l_{i}^{V} } \right)-1\right]\right\} \, .
\end{equation}

\begin{figure}[!ht]
\centering {
\begin{minipage}[!ht]{9.4 cm}
{\includegraphics[scale=0.8]{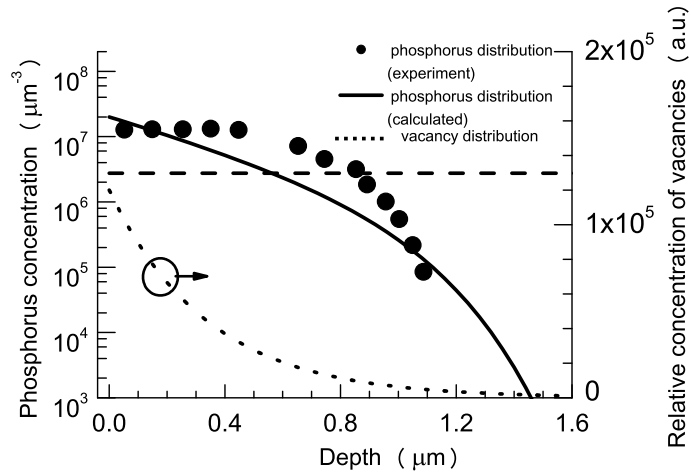}}
\end{minipage}
} \caption{Calculated phosphorus profile for radiation-enhanced
diffusion due to plasma treatment at a temperature of 820
${}^{o}$C. The dotted curve is the concentration of nonequilibrium
point defects normalized to the thermally equilibrium one. The
filled circles represent the experimental phosphorus profile from
\cite{Strack-63}. The dashed curve is the concentration of
intrinsic carriers . It is supposed that the diffusion of
phosphorus occurs due to the simple vacancy mechanism}
\label{fig:PlasmaVak}
\end{figure}

The phosphorus concentration profile calculated by means of
expression \eqref{PartialSolution} for silicon doping from the gas
discharge plasma is presented in Fig.~\ref{fig:PlasmaVak}. The
average migration length of point defects $l_{i}^{V} $ was chosen
to be equal to 0.34 $\mu $m.

As can be seen from Fig.~\ref{fig:PlasmaVak}, the impurity
concentration profile calculated from expression
\eqref{PartialSolution} disagrees qualitatively with the
experimental data describing the phosphorus radiation-enhanced
diffusion. Therefore, one can conclude that the simple vacancy
mechanism does not play an important role in the phosphorus
diffusion. On the other hand, the assumption that the phosphorus
diffusion occurs due to the ``impurity atom -- intrinsic point
defect'' pairs provides an excellent agreement between the
calculated and experimental concentration profiles for the
phosphorus radiation-enhanced diffusion \cite{Velichko-15}. The
agreement is observed in the entire diffusion zone including the
region near the semiconductor surface.

\subsection{Diffusion due to the formation, migration, and
dissociation of ``impurity atom -- intrinsic point defect'' pairs}

For simulation of the RED of phosphorus atoms within the framework
of the pair diffusion mechanism we use the experimental data of
\cite{Akutagawa-79} similar to investigations of
\cite{Velichko-97,Velichko-02}. However, in contrast to
\cite{Velichko-97} another set of the data was chosen, namely, the
phosphorus concentration profiles obtained for different values of
proton bombardment energy. To improve the fitting to the
experimental impurity profiles before and after the
radiation-enhanced diffusion, it was supposed that, due to the
impurity implantation and also to the proton bombardment,
relatively small amounts of phosphorus atoms occupy an
interstitial position and participate in the long-range
interstitial diffusion \cite{Velichko-12}. It is worth noting that
preliminary simulation shows that the average migration length of
point defects decreases with the proton energy. This phenomenon
can be explained if we suppose that the point defects responsible
for the RED interact with sinks located in the
phosphorus-implanted layer or annihilate with the nonequilibrium
point defects generated at the surface or in the layer damaged by
phosphorus implantation. We suppose that additional point defects
are generated because simulation shows that there is an additional
redistribution of implanted impurity directed to the region of
implanted protons. For example, if the radiation-enhanced
diffusion is governed by silicon self-interstitials, vacancies can
be these additional point defects according to the investigations
of \cite{Krause-Rehberg-00,Peeva-04}. We propose, for simplicity,
that these defect are generated on the surface, but this problem
requires a further investigation. With the assumptions made above,
the total flux of phosphorus is the sum of the E-center flux and
of the ``phosphorus atom -- silicon self-interstitial'' pairs
flux. To simulate this process, the diffusion equation obtained in
\cite{Velichko-84} can be used. In one dimension, Eq.
\eqref{DifEqLow} has the following form:

\begin{equation}\label{DifEqLow1D}
\frac{\partial \, C}{\partial \, t} =D_{i}^{E} \, \frac{\partial
^{\, 2} \left(\tilde{C}^{V\times } C\right)}{\partial \, x^{2} }
+D_{i}^{F} \, \frac{\partial ^{\, 2} \left(\tilde{C}^{I\times }
C\right)}{\partial \, x^{2} } \, .
\end{equation}

\begin{figure}[!ht]
\centering {
\begin{minipage}[!ht]{9.4 cm}
{\includegraphics[scale=0.8]{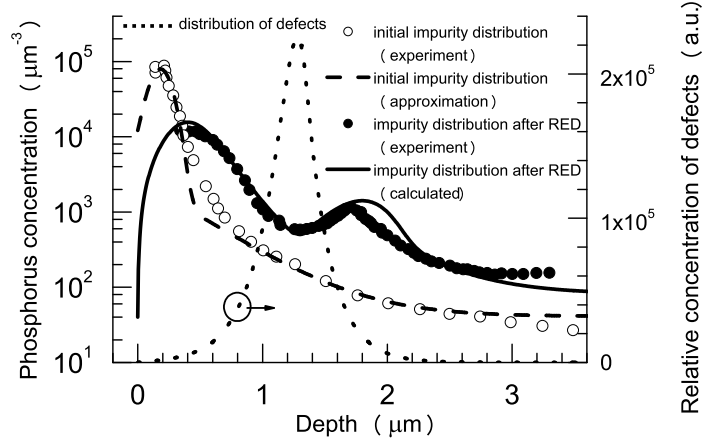}}
\end{minipage}
} \caption{Calculated phosphorus concentration profile for the
radiation-enhanced diffusion under conditions of proton
implantation with an energy of 140 keV at a temperature of 700
${}^{o}$C for 2 hours. The dashed and solid curves are
respectively the calculated distributions of phosphorus atoms
before and after diffusion, the open and filled circles are the
experimental phosphorus concentration profiles according to
\cite{Akutagawa-79}, the dotted curve represents the concentration
of nonequilibrium silicon self-interstitials normalized to the
thermally equilibrium one} \label{fig:Akutagawa140}
\end{figure}

\begin{figure}[!ht]
\centering {
\begin{minipage}[!ht]{9.4 cm}
{\includegraphics[scale=0.8]{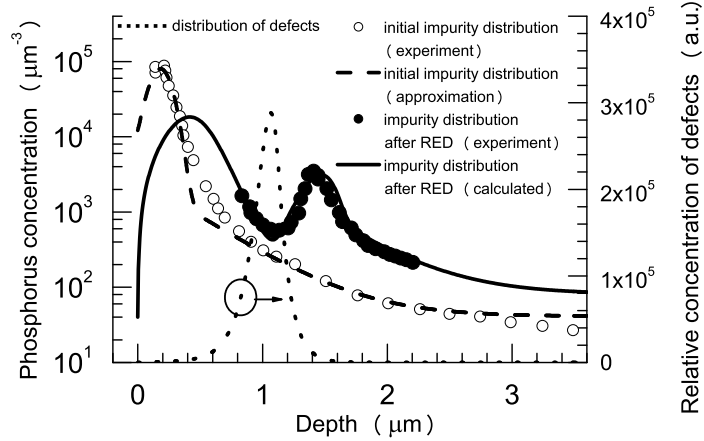}}
\end{minipage}
} \caption{Calculated phosphorus concentration profile for the
radiation-enhanced diffusion under conditions of proton
implantation with an energy of 110 keV at a temperature of 700
${}^{o}$C for 2 hours. The dashed and solid curves are the
calculated distributions of phosphorus atoms respectively before
and after diffusion, the open and filled circles are the
experimental phosphorus concentration profiles according to
\cite{Akutagawa-79}, the dotted curve represents the concentration
of nonequilibrium silicon self-interstitials normalized to the
thermally equilibrium one} \label{fig:Akutagawa110}
\end{figure}

\begin{figure}[!ht]
\centering {
\begin{minipage}[!ht]{9.4 cm}
{\includegraphics[scale=0.8]{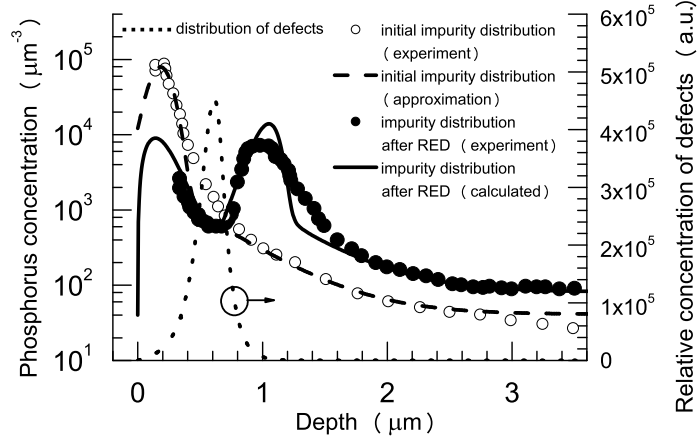}}
\end{minipage}
} \caption{Calculated phosphorus concentration profile for the
radiation-enhanced diffusion under conditions of proton
implantation with an energy of 80 keV at a temperature of 700
${}^{o}$C for 2 hours. The dashed and solid curves are the
calculated distributions of phosphorus atoms respectively before
and after diffusion, the open and filled circles are the
experimental phosphorus concentration profiles according to
\cite{Akutagawa-79}, the dotted curve represents the concentration
of nonequilibrium silicon self-interstitials normalized to the
thermally equilibrium one} \label{fig:Akutagawa80}
\end{figure}

The phosphorus concentration profiles calculated for the
radiation-enhanced diffusion owing its origin to the pair
diffusion mechanism described by Eq. \eqref{DifEqLow1D} are
presented in Figs.~\ref{fig:Akutagawa140},
~\ref{fig:Akutagawa110}, and ~\ref{fig:Akutagawa80}. The energy of
proton bombardment is equal to 140, 110, and 80 keV, respectively.
Proton bombardment was carried out at a temperature of 700
$^{\circ}$C for 2 hours at a beam density of 1.0 $\mu
$A/cm${}^{2}$.

To approximate the initial phosphorus concentration profile, we
used the value of the dose equal to 1.7$\times$10$^{12}$
ions/cm$^{2}$ that is 13.3 \% larger than the dose reported in
\cite{Akutagawa-79}. In addition, it is supposed that a small
fraction of implanted impurity (5.2$\times$10$^{2}$ ions/cm$^{2}$
occupied an interstitial position and participated in the
interstitial migration with an average migration length of 0.42
$\mu $m. Due to the long-range interstitial migration during
implantation and subsequent annealing, a ``tail'' region of the
initial phosphorus profile is formed. The interstitial phosphorus
atoms also form a ``tail'' region of the final impurity profile.
The dose of phosphorus interstitials generated during proton
bombardment is equal to 7.4$\times$10$^{2}$ ions/cm$^{2}$.

The following values of parameters that describe the implantation
of 140 keV hydrogen ions have been used for calculating the
phosphorus concentration profile presented in
Fig.~\ref{fig:Akutagawa140}: $R_{p} $ = 1.24 $\mu $m, $\Delta
R_{p} $ = 0.131 $\mu $m, $Sk$ = -5.44, $R_{m} $ = 1.31 $\mu $m
\cite{Burenkov-80}. Here $R_{p} $ and $\Delta R_{p} $ are the
average projective range of implanted ions and straggling of the
projective range, respectively, $Sk$ and $R_{m} $ are the skewness
and the position of the maximal value of the implanted ion
profile, respectively. It was supposed that the generation rate of
the nonequilibrium point defects responsible for the diffusion of
impurity atoms is proportional to the distribution of implanted
protons. The value of the intrinsic phosphorus diffusivity  = 4.
849 $\times$10${}^{-1}$${}^{0}$ $\mu $m${}^{2}$/s has been
calculated from the temperature dependence presented in
\cite{Haddara-00}. It is supposed that a fraction of intrinsic
diffusivity referred to as the diffusion due to the ``impurity
atom -- silicon self-interstitial'' pairs is equal to
4.0$\times$10${}^{-1}$${}^{0}$ $\mu $m${}^{2}$/s, whereas a
fraction related to the ``impurity atom -- vacancy'' pairs
(E-centers) is much smaller (0.849$\times$10${}^{-1}$${}^{0}$ $\mu
$m${}^{2}$/s). The Robin-type boundary condition providing the
removal of a part of phosphorus atoms was imposed on the
semiconductor surface.

The stationary distribution of nonequilibrium defects was obtained
from the numerical solution of the diffusion equation for point
defects and is presented in Fig.~\ref{fig:Akutagawa140} by a
dotted curve (silicon self-interstitials). The average migration
length of silicon self-interstitials in intrinsic silicon
$l_{i}^{I} =\sqrt{d_{i}^{I} \tau _{i}^{I} \, } $, derived from the
fitting of the calculated phosphorus concentration profile after
diffusion to the experimental one, is equal to 0.19 $\mu $m. Here
$d_{i}^{I} $ and $\tau _{i}^{I} $ are the diffusivity and average
lifetime of silicon self-interstitials in intrinsic silicon,
respectively. It follows from this fitting that the average
migration length of vacancies is equal to 0.3 $\mu $m if they are
generated on the semiconductor surface with a normalized surface
concentration equal to 1.7$\times$10${}^{5}$ a.u.

It can be seen from Fig.~\ref{fig:Akutagawa140} that there is a
good agreement between the calculated and experimental phosphorus
concentration profiles including the position of the local minimum
of impurity concentration and the form of ``uphill'' diffusion
region. Similar calculations for a proton energy of 110 keV are
presented in Fig.~\ref{fig:Akutagawa110}.

The following values of parameters that describe the implantation
of 110 keV hydrogen ions have been used for calculating the
phosphorus concentration profile presented in
Fig.~\ref{fig:Akutagawa110}: $R_{p}$ = 1.02 $\mu $m, $\Delta
R_{p}$ = 0.12 $\mu $m, $Sk$ = -4.73, and $R_{m}$ = 1.087 $\mu $m.
It is worth noting that the values of $\Delta R_{p} $ and $Sk$ are
taken from \cite{Burenkov-80}. On the other hand, the value
$R_{p}$ = 1.02 $\mu $m is greater than $R_{p}$ = 0.982 $\mu $m
from \cite{Burenkov-80}. Using the value $R_{p}$ = 0.982 $\mu $m
results in a more poor agreement with experimental data,
especially for the position of the local minimum on the phosphorus
concentration profile. The average migration length of silicon
self-interstitials $l_{i}^{I} $, derived from the fitting to
experimental data, is equal to 0.09 $\mu $m, i. e., less than the
average migration length for bombardment with a proton energy of
140 keV. The dose of phosphorus interstitials generated during
proton bombardment and participated in the long-range migration is
equal to 9.2$\times $10$^{2}$ ions/cm$^{2}$. The other modeling
parameters are the same as in the previous case of 140 keV proton
bombardment.

Finally, the results of modeling of the radiation-enhanced
diffusion for a proton energy of 80 keV are presented in
Fig.~\ref{fig:Akutagawa80}.

The following values of parameters that describe the implantation
of 80 keV hydrogen ions have been used for calculating the
phosphorus concentration profile presented in
Fig.~\ref{fig:Akutagawa80}: $R_{p}$ = 0.572 $\mu $m, $\Delta
R_{p}$ = 0.1092 $\mu $m, $Sk$ = -3.96, and $R_{m} $ = 0.6304 $\mu
$m. The average migration length of silicon self-interstitials
$l_{i}^{I} $, derived from the fitting to experimental data, is
equal to 0.08 $\mu $m, i. e., practically it equals the average
migration length for the case of bombardment with a proton energy
of 110 keV. The dose of phosphorus interstitials generated during
proton bombardment and participated in the long-range migration is
equal to 7.4$\times $10$^{2}$ ions/cm$^{2}$. The vacancies are
generated at the semiconductor surface with a normalized surface
concentration equal to 5.0$\times$10$^{4}$ a.u. The other modeling
parameters are the same as in the previous cases of proton
bombardment with energies of 140 and 110 keV.

It is worth noting that $\Delta R_{p}$ = 0.1092 $\mu $m and $Sk$ =
-3.96 are taken from \cite{Burenkov-80}. On the other hand, the
used value $R_{p}$ = 0.572 $\mu $m is less than $R_{p}$ = 0.742
$\mu $m from \cite{Burenkov-80} or $R_{p}$ = 0.7499 $\mu $m from
\cite{Burenkov-85}. It is interesting to note that $R_{p} $ =
0.572 $\mu $m, derived from the fitting procedure, is
approximately equal to $R_{pD}$ = 0.611 $\mu $m
\cite{Burenkov-85}, where $R_{pD} $ is the position of the maximum
for distribution of the energy deposited in atomic collisions in
silicon. However, taking into account the different sign of
deviation between adjustable and tabulated values of $R_{p} $ for
proton energies of 80 and 110 keV, we suppose that, perhaps, the
reason was the inaccuracy of the measurements of the phosphorus
concentration profiles in \cite{Akutagawa-79}. The deviation of
derived and tabulated $R_{p} $ also can occur due to the more
complicated interaction between generated defects and previously
formed defects or phosphorus atoms in the implanted layer. Further
investigations of the RED on the basis of precise SIMS
measurements are required to solve this interesting problem. In
any case, the good agreement between the calculated and
experimental phosphorus concentration profiles for different
energies of proton bombardment confirms the conclusion of
\cite{Velichko-97} that the low-temperature radiation-enhanced
diffusion of phosphorus atoms occurs due to the formation,
migration, and dissociation of ``impurity atom -- point defect''
pairs. Relying on this knowledge, we can investigate the character
form of impurity concentration profiles that can be formed due to
the radiation-enhanced diffusion in the layered semiconductor
structures.

\section{Simulation of the radiation-enhanced diffusion in thin
layers}

\begin{figure}[!ht]
\centering {
\begin{minipage}[!ht]{9.4 cm}
{\includegraphics[scale=0.8]{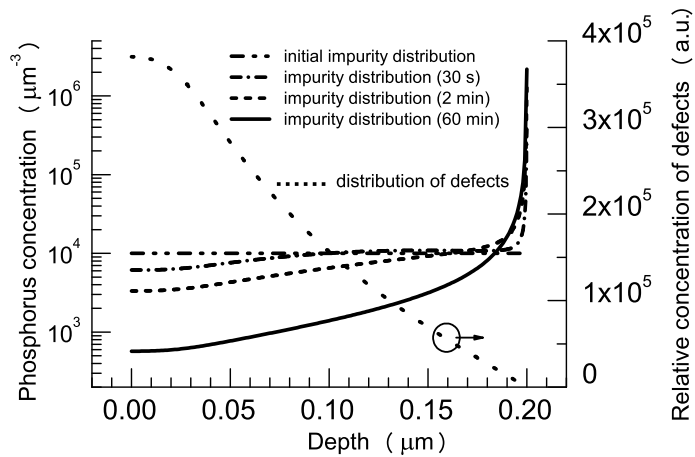}}
\end{minipage}
} \caption{Calculated phosphorus concentration profiles for the
radiation-enhanced diffusion under conditions of proton
bombardment with an energy of 1 keV at a temperature of 700
${}^{o}$C for 0.5, 2, and 60 minutes. The dash-double-dotted curve
represents the initial impurity distribution, the dash-dotted,
dashed, and solid curves are the calculated distributions of
phosphorus atoms after the radiation-enhanced diffusion for 0.5,
2, and 60 minutes, respectively, the dotted curve represents the
concentration of nonequilibrium silicon self-interstitials
normalized to the thermally equilibrium one} \label{fig:Depletion}
\end{figure}

Let us consider a 0.2-$\mu $m thick uniformly doped silicon layer
formed by silicon on insulator technology. Due to the small
thickness of the layer, 1 keV proton implantation from the remote
hydrogen containing plasma can be used to provide
radiation-enhanced diffusion. If the silicon substrate temperature
is equal to 700 ${}^{o}$C, there is no passivation of impurity by
hydrogen atoms \cite{Chevallier-02}. The results of modeling of
the radiation-enhanced diffusion phosphorus atoms for 0.5, 2, and
60 minutes are presented in Fig.~\ref{fig:Depletion}.

The following values of parameters that describe the implantation
of 1 keV hydrogen ions have been used for calculating the
phosphorus concentration profiles after the RED presented in
Fig.~\ref{fig:Depletion}: $R_{p}$ = 0.01631 $\mu $m, $\Delta
R_{p}$ = 0.015 $\mu $m, $Sk$ = -0.3, and $R_{m}$ = 0.0184 $\mu $m
\cite{Burenkov-85}. The average migration length of silicon
self-interstitials $l_{i}^{I} $ has been chosen to be equal to
0.09 $\mu $m. The conditions for the generation of silicon
self-interstitials are similar to the previously investigated
cases of phosphorus radiation-enhanced diffusion. For simplicity,
reflecting boundary conditions have been chosen for impurity atoms
on the interface between the silicon layer and insulator and on
the semiconductor surface. It is supposed that the reflecting
boundary condition is also valid for the defect diffusion near the
semiconductor surface whereas the boundary condition describing
the absorption of silicon self-interstitials is used at the
interface between the silicon layer and insulator.

\begin{figure}[!ht]
\centering {
\begin{minipage}[!ht]{9.4 cm}
{\includegraphics[scale=0.8]{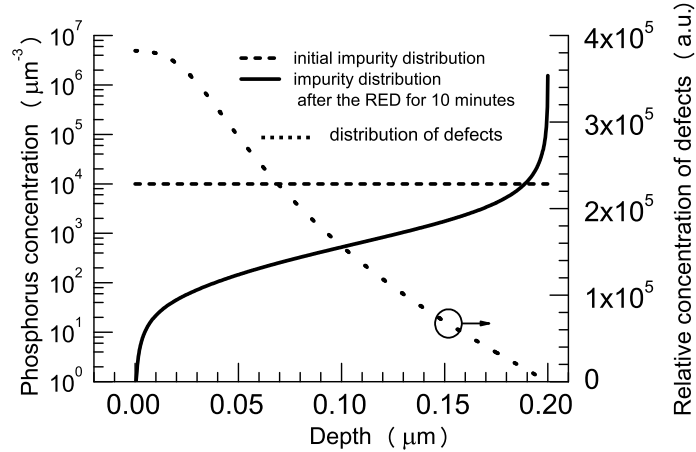}}
\end{minipage}
} \caption{Calculated phosphorus concentration profiles for the
radiation-enhanced diffusion under conditions of proton
bombardment with an energy of 1 keV at a temperature of 700
${}^{o}$C for 10 minutes. The dashed and solid curves represent
the initial impurity distribution and the calculated distribution
of phosphorus atoms after the radiation-enhanced diffusion,
respectively, the dotted curve represents the concentration of
nonequilibrium silicon self-interstitials normalized to the
thermally equilibrium one. It is supposed that evaporation of
phosphorus atoms from the surface occurs during plasma treatment}
\label{fig:DepletionRobin}
\end{figure}

As can be seen from Fig.~\ref{fig:Depletion}, depletion in
impurity concentration occurs practically within the whole layer
due to the RED. Simultaneously, a narrow region with a high
impurity concentration is formed in the vicinity of the interface
between the silicon layer and insulator. In
Fig.~\ref{fig:DepletionRobin} a similar calculation for the
Robin-type boundary condition imposed on the semiconductor surface
is presented. It is supposed that the duration of the RED is equal
to 10 minutes and impurity atoms can evaporate from the surface.
It can be seen from Fig.~\ref{fig:DepletionRobin} that for this
case of the radiation-enhanced diffusion an additional decrease in
the impurity concentration occurs in the near surface region due
to evaporation of impurity atoms.

Taking into account that strictly assigned local areas of the
surface can be exposed to ion bombardment, we can form the local
domains with depleted impurity concentration in the bulk and with
a high concentration of impurity atoms on the boundary of these
selected domains. Due to the low temperature of the
radiation-enhanced diffusion, the redistribution of impurity atoms
in other regions of a semiconductor will be negligible. This
phenomenon can be useful for the technology of semiconductor
devices and integrated microcircuits.

\section{Conclusions}

To investigate the microscopic mechanisms of impurity transport in
semiconductors, modeling of phosphorus radiation-enhanced
diffusion during implantation of high-energy protons to a silicon
substrate being at an elevated temperature and during the
treatment of a silicon substrate in the hydrogen-containing plasma
with addition of a diffusant has been made. It follows from the
comparison of the calculated impurity profiles with experimental
ones that the radiation-enhanced diffusion occurs due to the
formation, migration, and dissociation of the ``impurity atom --
intrinsic point defect'' pairs which are in a local thermodynamic
equilibrium with the substitutionally dissolved impurity atoms and
nonequilibrium point defects generated as a result of external
irradiation. For a proton energy of 140 keV the position of the
maximum of defect generation rate is equal to the maximum
concentration of implanted hydrogen. This is the evidence in favor
of the statement that silicon self-interstitials are the defects
responsible for the phosphorus radiation-enhanced diffusion. If
the proton energy decreases, the derived value of the average
migration length of nonequilibrium silicon self-interstitials also
decreases from 0.19 $\mu $m for an energy of 140 keV to 0.09 and
0.08 $\mu $m for energies of 110 and 80 keV, respectively. The
decrease of the average migration length can occur due to the
interaction of silicon self-interstitials with the vacancies
generated at the surface or with the defects formed in the
phosphorus implanted region. Indeed, as follows from the
experimental data \cite{Akutagawa-79}, the position of the local
minimum of phosphorus concentration is shifted to the surface with
proton energy decrease. This shift is a natural consequence of the
decrease in the distance between the surface and position of
maximum of defect generation. On the other hand, the position of
the maximum of defect generation and simultaneously the maximum of
the implanted hydrogen concentration $R_{m} $ = 0.6304 $\mu $m
obtained by fitting the experimental phosphorus profile for a
proton implantation energy of 80 keV are less than $R_{m}$ = 0.8
$\mu $m calculated from the tables in \cite{Burenkov-80}. We
suppose that, perhaps, there was the inaccuracy of measurements of
phosphorus concentration profiles in \cite{Akutagawa-79}. The
deviation of the derived and tabulated $R_{m} $ can also occur due
to the more complicated interaction between the generated defects
and previously formed defects or phosphorus atoms in the implanted
layer. Further investigations of the RED on the basis of precise
SIMS measurements are required to solve this interesting problem.

It is worth noting that calculation for the case of simple vacancy
mechanism of diffusion due to the exchange of places between the
impurity atom and neighboring vacancy disagrees qualitatively with
the experimental data of \cite{Strack-63} for the phosphorus
radiation-enhanced diffusion occurring during plasma treatment. To
provide the latter calculation, an analytical solution of the
diffusion equation that describes the quasistationary diffusion of
impurity atoms under condition of the sputtering of a surface was
obtained.

On the basis of the pair diffusion mechanism, a theoretical
investigation was carried out for the form of impurity profiles
that can be created in thin silicon layers due to the
radiation-enhanced diffusion. It was shown that depletion of the
uniformly doped silicon layer occurs during plasma treatment
except for the vicinity of the interface between the silicon and
insulator where a narrow region with a high impurity concentration
is formed. These characteristic features of the dopant profile are
due to the nonuniform distribution of point defects responsible
for the impurity diffusion.

The results of the calculations performed give a clear evidence in
favor of conducting further investigation of various doping
processes based on the radiation-enhanced diffusion, especially
the processes of plasma doping, to develop a cheap method for the
formation of special impurity distributions in the local
semiconductor domain.

\end{document}